# EINSTEIN'S COEFFICIENTS AND THE NATURE OF THERMAL RADIO EMISSION

Fedor V.Prigara

*Institute of Microelectronics and Informatics , Russian Academy of Sciences ,*
*21 Universitetskaya , 150007 Yaroslavl , Russia*

The relations between Einstein's coefficients for spontaneous and induced emission of radiation with account for the natural linewidth are obtained . It is shown that thermal radio emission is stimulated one . Thermal radio emission of non-uniform gas is considered .

The analysis of observational data on thermal radio emission from various astrophysical objects leads to the condition of emission implying the induced emission of radiation [5]. Here we show that the stimulated character of thermal radio emission follows from the relations between Einstein's coefficients for spontaneous and induced emission of radiation .

Consider two level system (atom or molecule ) with energy levels $E_1$ and $E_2 > E_1$ which is in equilibrium with thermal blackbody radiation . We denote as $\nu_s = A_{21}$ the number of transitions from the upper energy level to the lower one per unit time caused by a spontaneous emission of radiation with the frequency $\omega = (E_2 - E_1)/\hbar$, where $\hbar$ is the Planck constant . The number of transitions from the energy level $E_2$ to the energy level $E_1$ per unit time caused by a stimulated emission of radiation may be written in the form

$$\nu_i = B_{21} B_\nu \Delta\nu, \qquad (1)$$

where

$$B_\nu = \hbar\omega^3 / (2\pi^2 c^2)(\exp(\hbar\omega/kT) - 1) \qquad (2)$$

is a blackbody emissivity (Planck's function ) , $T$ is the temperature , $k$ is the Boltzmann constant , $c$ is the speed of light , and $\Delta\nu$ is the linewidth .

The number of transitions from the lower level to the upper one per unit time may be written in the form

$$\nu_{12} = B_{12} B_\nu \Delta\nu, \qquad (3)$$

Here $A_{21}, B_{21}$ and $B_{12}$ are the coefficients introduced by Einstein [1,2], the coefficients $B_{12}$ and $B_{21}$ being modified to account for the linewidth $\Delta\nu$.

We denote as $N_1$ and $N_2$ the number of atoms occupying the energy levels $E_1$ and $E_2$, respectively. The levels $E_1$ and $E_2$ are suggested for simplicity to be non-degenerated.

In the equilibrium state the full number of transitions from the lower level to the upper one is equal to the number of reverse transitions :

$$N_1 \nu_{12} = N_1 B_{12} B_\nu \Delta\nu = N_2 \nu_{21} = N_2 (A_{21} + B_{21} B_\nu \Delta\nu). \qquad (4)$$

From the last equation we obtain

$$N_2 / N_1 = B_{12} B_\nu \Delta\nu / (A_{21} + B_{21} B_\nu \Delta\nu). \qquad (5)$$

The linewidth $\Delta\nu$ is suggested to be equal to the natural linewidth [3,4] :

$$\Delta\nu = A_{21} + B_{21} B_\nu \Delta\nu. \qquad (6)$$

It follows from the last equation that

$$\Delta\nu = A_{21} / (1 - B_{21} B_\nu). \qquad (7)$$

Substituting this expression in the equation (5), we obtain

$$N_2 / N_1 = B_{12} B_\nu. \qquad (8)$$

In the limit of higher temperatures $T \to \infty$, corresponding to the range of frequencies $\hbar\omega < kT$, the function $B_\nu(T)$ is given by the Rayleigh-Jeans formula

$$B_\nu = 2kT\nu^2 / c^2, \qquad (9)$$

where $\nu$ is the frequency of radiation, $\nu = \omega / 2\pi$.

Since $B_\nu \to \infty$ when $T \to \infty$, it follows from the equation (7) that the coefficient $B_{21}$ is depending on the temperature $T$ in such a manner that $B_{21} B_\nu < 1$. It is clear that $B_{21} B_\nu \to 1$ when $T \to \infty$. The ratio of frequencies of transitions caused by spontaneous and induced emission of radiation is given by the expression

$$\nu_s / \nu_i = A_{21} / (B_{21} B_\nu \Delta\nu) = (1 - B_{21} B_\nu) / (B_{21} B_\nu). \qquad (10)$$

This ratio is approaching zero when $T \to \infty$, since $B_{21} B_\nu \to 1$. It means that in the range of frequencies $\hbar\omega < kT$ thermal radiation is produced by the stimulated emission, whereas the contribution of a spontaneous emission may be neglected.

The quantum of radio emission has the energy $\hbar\omega < 10^{-3} eV$. In the same time the temperature of radio emitting gas in the case of astrophysical objects usually exceeds $10^2$ K, so the condition $\hbar\omega < kT$ is valid. Thus thermal radio emission of gas nebulae and other astrophysical objects has the induced origin.

Thermal radio emission of non-uniform gas is produced by the ensemble of individual emitters. Each of these emitters is a molecular resonator the size of which has an order of magnitude of mean free path of photons

$$l = \frac{1}{n\sigma} \qquad (11),$$

where $n$ is the number density of particles of gas, and $\sigma$ is the photoabsorption cross-section.

The photoabsorption cross-section has an order of magnitude of atomic cross-section $\sigma=10^{-15}$ cm$^2$. Because of a stimulated emission, the attenuation of radio emission will be sufficiently small, although the value of photoabsorption cross-section is relatively great. In the range of frequencies $\hbar\omega < kT$ $B_{12}B_\nu = N_2/N_1 \approx 1$ and therefore $B_{12}B_\nu \approx B_{21}B_\nu$. It means that the number of transitions from the lower energy level to the upper one caused by the absorption of photon will be approximately equal to the number of reverse transitions from the upper level to the lower one caused by the stimulated emission of quantum of the same character as the original one.

Note that owing to the induced absorption of light described above Kirchhoff's law [1] is not valid in the range of frequencies $\hbar\omega < kT$, where the induced emission of radiation dominates and $B_{12}B_\nu \approx B_{21}B_\nu$.

The emission of each molecular resonator is coherent, with the wavelength
$$\lambda = l, \qquad (12)$$
and thermal radio emission of gaseous layer is incoherent sum of radiation produced by individual emitters.

The condition (12) implies that the radiation with the wavelength $\lambda$ is produced by the gaseous layer with the definite number density of particles $n$.

In the gaseous disk model, describing radio emitting gas nebulae [5], the number density of particles decreases reciprocally with respect to the distance $r$ from the energy center
$$n \propto r^{-1}. \qquad (13)$$

Together with the condition of emission (12) the last equation leads to the wavelength dependence of radio source size :

$$r_\lambda \propto \lambda . \qquad (14)$$

The relation (14) is indeed observed for sufficiently extended radio sources [6,7,8]. For instance , the size of radio core of galaxy M31 is 3.5 arcmin at the frequency 408 MHz and 1 arcmin at the frequency 1407 MHz [8] .

In the case of some compact radio sources instead of the relation (14) the relation

$$r_\lambda \propto \lambda^2 \qquad (15)$$

is observed [9] . This relation may be explained by the effect of strong magnetic field on the distribution of ionized gas density which changes the equation (13) for the equation

$$n \propto r^{-1/2} . \qquad (16)$$

In the limit of very strong magnetic fields the density of ionized gas does not depend on the radius $r$ .

The spectral density of flux from an extended radio source is given by the formula

$$F_\nu = \frac{1}{a^2} \int_0^{r_\lambda} B_\nu(T) \times 2\pi r dr \ , \qquad (17)$$

where $a$ is a distance from radio source to the detector of radiation .

The extended radio sources may be divided in two classes . Type 1 radio sources are characterized by a stationary convection in the gaseous disk with an approximately uniform distribution of the temperature $T \approx const$ giving the spectrum

$$F_\nu \approx const . \qquad (18)$$

Type 2 radio sources are characterized by outflows of gas with an approximately uniform distribution of gas pressure $P=nkT \approx const$ . In this case the equation (13) gives

$$T \propto r , \qquad (19)$$

so the radio spectrum , according to the equation (17) , has the form

$$F_\nu \propto \nu^{-1} . \qquad (20)$$

Both classes include numerous galactic and extragalactic objects . In particular , edge-brightened supernova remnants [10] belong to the type 2 radio sources in accordance with the relation (19) , whereas center-brightened supernova remnants belong to the type 1 radio sources . Steep spectrum radio quasars [15] belong to type 2 radio sources , and flat spectrum radio quasars [15] belong to type 1 radio sources .

There is a direct observational evidence that the outflows of gas have radio spectra describing by the equation (20) . This evidence was obtained for the bright radio condensations emerging in opposite directions from a compact core of the radio counterpart of GRS1915+105 [11] .

Note that the synchrotron inerpretation of radio emission from various astrophysical objects [16] encounters the essential difficulties . The synchrotron theory is unable to explain the correlation between the spectral index and the radial distribution of brightness for supernova remnants [10], as well as the wavelength dependence of radio source size , since a spectral index according to the synchrotron theory has a local origin [16] . The constancy of the spectral index for a jet [17] also cannot be explained by the synchrotron theory predicting the evolution of a radio spectrum to the steeper one .

Planetary atmospheres are similar to type 1 radio sources , since the temperature of radio emitting atmosphere layers is approximately constant , whereas the gas pressure is exponentially decreasing with the increase of height . Since the total thickness of radio emitting layers is small with respect to the radius of a planet , the spectrum of thermal radio emission of planetary atmosphere is similar to the radio spectra of planetary nebulae in the long wavelengths range [12] . Notice that at the higher frequencies radio emission of planetary nebulae is described by the law (18) .

The induced origin of thermal radio emission is consistent with the existence of maser sources associated with gas nebulae and nuclei of galaxies [13,14] .

The higher brightness temperatures of compact extragalactic radio sources [6] are explained by maser amplification of thermal radio emission . This conclusion is supported by the constancy of the brightness temperature of the radio core of quasar NRAO 530 in the wavelength range of 0.3 cm to 2 cm , the flux density obeying the law (18) in the wavelength range of 0.35 cm to 6.25 cm [6] . It is quite possible that the higher brightness temperatures of radio pulsars are also connected with maser amplification of thermal radio emission .

It is worthwhile to notice that the original Einstein's theory [1] which does not take into account a linewidth also predicts the induced character of thermal radio emission . However the account for the natural linewidth essentially affects the form of relation (8) .

In the limit of $T \to 0$ , corresponding to the range of frequencies $\hbar\omega > kT$ , the function $B_\nu(T)$ is given by the formula

$$B_\nu = (\hbar\omega^3 / 2\pi^2 c^2) \exp(-\hbar\omega / kT) \qquad (21)$$

and the relation (8) gives Boltzmann's law

$$(N_2 / N_1) = \exp(-\hbar\omega / kT) \qquad (22)$$

if we take

$$B_{12} = 2\pi^2 c^2 / \hbar\omega^3 . \qquad (23)$$

If we suggest in addition that $B_{12} = B_{21}$ , then the linewidth will be given by the formula

$$\Delta\nu = A_{21} / (1 - \exp(-\hbar\omega / kT)) . \qquad (24)$$

The ratio of frequencies of transitions caused by induced and spontaneous emission of radiation respectively is given by the expression

$$\nu_i / \nu_s = B_{21} B_\nu / (1 - B_{21} B_\nu) \approx \exp(-\hbar\omega / kT) . \qquad (25)$$

In this range of frequencies the contribution of spontaneous emission dominates .

Notice that the relation $B_{12} = B_{21}$ is no more than hypothesis which cannot be justified by thermodynamical approach . In fact this hypothesis might not be valid and then the ratio $\nu_i / \nu_s$ may exceed the unit. It is quite possible since the coefficients $B_{12}$, and $B_{21}$ are depending on the temperature.

Thus it follows from the relations between Einstein's coefficients that thermal radio emission has a stimulated character. The induced origin of thermal radio emission leads to the condition of emission (12) describing thermal radio emission of non-uniform gas.

The author is grateful to A.V.Postnikov and V.V.Ovcharov for useful discussions.

---

[1] M.Jammer, *The Conceptual Development of Quantum Mechanics* (McGraw-Hill, New York, 1967).

[2] J.R.Singer, *Masers* ( Wiley, New York, 1959 ), Ch.II.

[3] L.D.Landau, and E.M.Lifshitz, *Quantum Mechanics, Non-Relativistic Theory*

( Addison-Wesley, Reading, MA, 1958 ).

[4] V.B.Berestetskii, E.M.Lifshitz, and L.P.Pitaevskii, *Quantum Electrodynamics* (Nauka, Moscow, 1989 ).

[5] F.V.Prigara, astro-ph/0110399 (2001).

[6] G.C.Bower, and D.C.Backer. Astrophys. J. **507**, L117 (1998).

[7] C.Tadhunter *et al.* Mon. Not. R. Astron. Soc. **327**, 227 (2001).

[8] A.S.Sharov, *The Andromeda Nebula* ( Nauka, Moscow, 1982 ).

[9] K.Y.Lo *et al.,* Nature (London) **361**, 38 (1993). K.Y.Lo, in *AIP Proc. 83 : The Galactic Center*, edited by G.Riegler, and R.Blandford ( AIP, New York, 1982 ).

[10] S.R.Kulkarni, and D.A.Frail, Nature ( London ) **365**, 33 (1993).

[11] I.F.Mirabel, and L.F.Rodriguez. Nature (London) **371**, 46 (1994).

[12] S.R.Pottasch , *Planetary Nebulae* (D.Reidel , Dordrecht , 1984 ) .

[13] M.Miyoshi *et al.,* Nature ( London ) **373** , 127 (1995) .

[14]  N.G.Bochkarev , *Basic Physics of Interstellar Medium* (Moscow University Press , Moscow , 1992 ) .

[15] T.A.Rector, and J.T.Stocke , astro-ph/0105100 (2001) .

[16] K.R.Lang , *Astrophysical Formulae* ( Springer-Verlag , Berlin , 1974 ) .

[17] V.G.Gorbatsky , *An Introduction to  Physics of Galaxies and Their Clusters* (Nauka, Moscow, 1986 ) .